\documentclass[reprint, superscriptaddress,amsmath,amssymb,aps]{revtex4-2}
\usepackage[utf8]{inputenc}

\usepackage{graphicx}
\usepackage{physics}
\usepackage{dcolumn}
\usepackage{bm}
\usepackage[dvipsnames]{xcolor}
\usepackage{hyperref}
\usepackage[mathlines]{lineno}

\begin{document}

\preprint{APS/123-QED}

\title{Spin-orbit coupling of optical vector vortices in coherently prepared media}

\author{Dharma P. Permana}
\email{dharma.permana@ff.stud.vu.lt}
\email{dhp.permana@gmail.com}
\affiliation{%
 Institute of Theoretical Physics and Astronomy, Vilnius University, Saul\.{e}tekio 3, Vilnius LT-10257, Lithuania
}%
\affiliation{Laser Research Center, Vilnius University, Vilnius, LT-10223, Lithuania}
\affiliation{D\'{e}partement Physique, Facult\'{e} des Sciences, Aix-Marseille Universit\'{e}, Marseille, France}

\author{Mazena Mackoit Sinkevi\v{c}ien\.{e}}%
 \email{mazena.mackoit-sinkeviciene@ff.vu.lt}
\affiliation{%
 Institute of Theoretical Physics and Astronomy, Vilnius University, Saul\.{e}tekio 3, Vilnius LT-10257, Lithuania
}%

\author{Julius Ruseckas}
\email{julius.ruseckas@gmail.com}
\affiliation{Baltic Institute of Advanced Technology, LT-01403 Vilnius, Lithuania}
\author{Hamid R. Hamedi}
 \email{hamid.hamedi@tfai.vu.lt}
\affiliation{%
 Institute of Theoretical Physics and Astronomy, Vilnius University, Saul\.{e}tekio 3, Vilnius LT-10257, Lithuania
}%



\date{\today}

\begin{abstract}
We investigate the propagation of an optical vector vortex weakly interacting with a coherently prepared atomic medium (phaseonium) in a three-level $\Lambda$ configuration. The vector beam consists of vortex pulse pairs with right- and left-circular polarizations, corresponding to opposite spin angular momentum (SAM), and carrying opposite orbital angular momentum (OAM) charges $\pm l$.
We show that during the propagation of the vortex pairs, analytically obtained in the linear regime, the medium inherits the topology of the vortex pair, mapping the OAM onto a spatially structured atomic coherence. This mapping produces $2|l|$-fold azimuthal transparency structures that reshape the beam intensity from a ring into a petal-like pattern. The OAM-structured atomic coherence induces a corresponding optical anisotropy within the medium, which feeds back into the propagating vector beam, resulting in optical spin–orbit coupling manifested as SAM exchange, rotation, and evolution of polarization textures. Depending on the initial ground-state population of the phaseonium, the polarization state evolves between left-circular, linear, and right-circular polarizations.  
\end{abstract}

\maketitle


\section{\label{Introduction}Introduction}
Over the past decades, optical vortices has been extensively studied since their first introduction by Coullet et al. \cite{coullet1989}. Optical vortex itself is a special type of light beam carrying orbital angular momentum (OAM) with discrete value of $l\hbar$, where $l$ can take any integer value \cite{allen1992}. The integer $l$ is directly related to the topological charge of the phase singularity, which corresponds to the helical phase factor $e^{il\phi}$, where $\phi$ is the azimuthal coordinate in the plane transverse to the propagation direction \cite{brambilla1991}. Combined with the photon spin angular momentum (SAM), which is associated with the beam’s polarization \cite{zhang2010}, optical vortex has become a key research area for quantum communication and information processing in higher dimensions \cite{gibson2004,park2018,erhard2018,plachta2022}.

Many important effects arise when optical fields interact with atomic ensembles, opening up a wide range of possibilities for controlling information processing using optical vortices. Notably, the effects arising from the coherent superpositions of atomic states have led to the discovery of electromagnetically induced transparency (EIT) \cite{boller1991,fleischhauer2000,Marangos.RMP2005}, slow light propagation \cite{paspalakis2002,juzeliunas2004}, enhancement of optical nonlinearities \cite{minxiao2001,gong2004,hamedi2015}, and formations of adiabatons \cite{manka1996,kudriasov2025adiabatons}. In particular, for EIT, a transparency window can form in an otherwise opaque medium due to quantum coherence effects induced by a control field, leading to modification of the medium susceptibility \cite{paspalakis2002,Lukin.Nature2001,Finkelstein2023}. The transparency window in EIT has been shown to depend on the properties of both the medium and the light \cite{boller1991,tsai2010}, as well as by the presence of an external magnetic field \cite{jelenkovic2008,vijayan2010,moon2010,wilson-gordon2013}.

The rich variety of new phenomena demonstrated by EIT-enabled atomic systems, together with the unique properties of optical vortices, has sparked extensive investigations into their interactions. These studies have led to the demonstration of exotic effects such as the transfer of OAM of light \cite{ruseckas2013,Hamid.PRA2019}, azimuthally dependent transparency \cite{hamid2018}, and entanglement of OAM state in photon-pairs \cite{guo2008}. However, most of these works have focused on cases where the vortex beam is uniformly polarized (a scalar vortex), leaving the manipulation of the SAM of vortex beams largely unexplored.

Contrary to scalar vortex beams, an optical vector vortex beam allows utilization of the SAM of the beam to achieve a wider range of applications. A vector vortex beam is composed of a superposition of two beams with orthogonal polarizations, resulting in a spatially varying polarization distribution \cite{rosales2018}. Such a beam can be described as a vector superposition of Laguerre–Gaussian (LG) modes carrying OAM with right- and left-circularly polarized (RCP and LCP) components \cite{yao2011}. The resulting spatially varying polarization can exhibit radial, azimuthal, or spiral patterns when the superposed LG modes have equal but opposite topological charges $l$ \cite{Yao2018,Tarak.OE2022,ma2025}.

Recently, studies of vector vortex interactions with EIT-enabled atomic media have revealed several intriguing phenomena. Among them is spatially dependent transparency \cite{radwell2017, tarak2017}, in a four-level tripod atomic system, which can be utilized as an atomic compass \cite{castellucci2021}. The propagation of vector vortex beams and transfer of SAMs in such atomic media have also been explored \cite{Kudriasov.OE2025}. In addition, a different four-level system in a V configuration has been employed to demonstrate the transfer of structured polarization \cite{Li.OE2022}. Yet, these schemes rely on the presence of a weak transverse magnetic field to establish a closed coupling loop among the atomic states and to facilitate coherent superposition. Consequently, their applicability is limited to systems in which a weak transverse magnetic field is present.

In this work, we demonstrate that the SAM exchange and spatially dependent transparency effects, previously observed in four-level tripod systems under external magnetic fields, can also be achieved without any magnetic field, relying solely on atomic coherence in a weakly interacting, coherently prepared medium. We investigate the propagation of an optical vector vortex, composed of a pair of vortex pulses with orthogonal circular polarizations and opposite OAM charges, through an ensemble of atoms in a three-level $\Lambda$-type configuration. The atoms are initially prepared in a ground-state superposition (phaseonium) \cite{scully1992}. By deriving the steady-state solution for the atomic coherences and analytically solving the Maxwell–Bloch equations in the linear regime (neglecting diffraction effects), we find that the atomic medium fully inherits the vector topology of the vortex. Specifically, the OAM is mapped onto the spatial atomic coherence,  giving rise to azimuthally dependent transparency windows with a $2|l|$-fold degeneracy that preserve their topological structure during vortex propagation. This spatially structured coherence induces an optical anisotropy in the medium, which channels back into the propagating vector vortex, enabling polarization conversion, rotation, and modification of the spin texture. In this sense, the propagation dynamics realize a coherence-induced optical spin–orbit coupling. Overall, this work establishes a clean interface for transferring the vector topology of light onto matter and for controlling coupled SAM–OAM dynamics through atomic coherence.

\section{\label{Theoretical Model}Theoretical Model}
\subsection{\label{Atom-light configuration}Atom–Light Interaction Hamiltonian}
\begin{figure}[!h]
    \centering
    \includegraphics[width=1\linewidth]{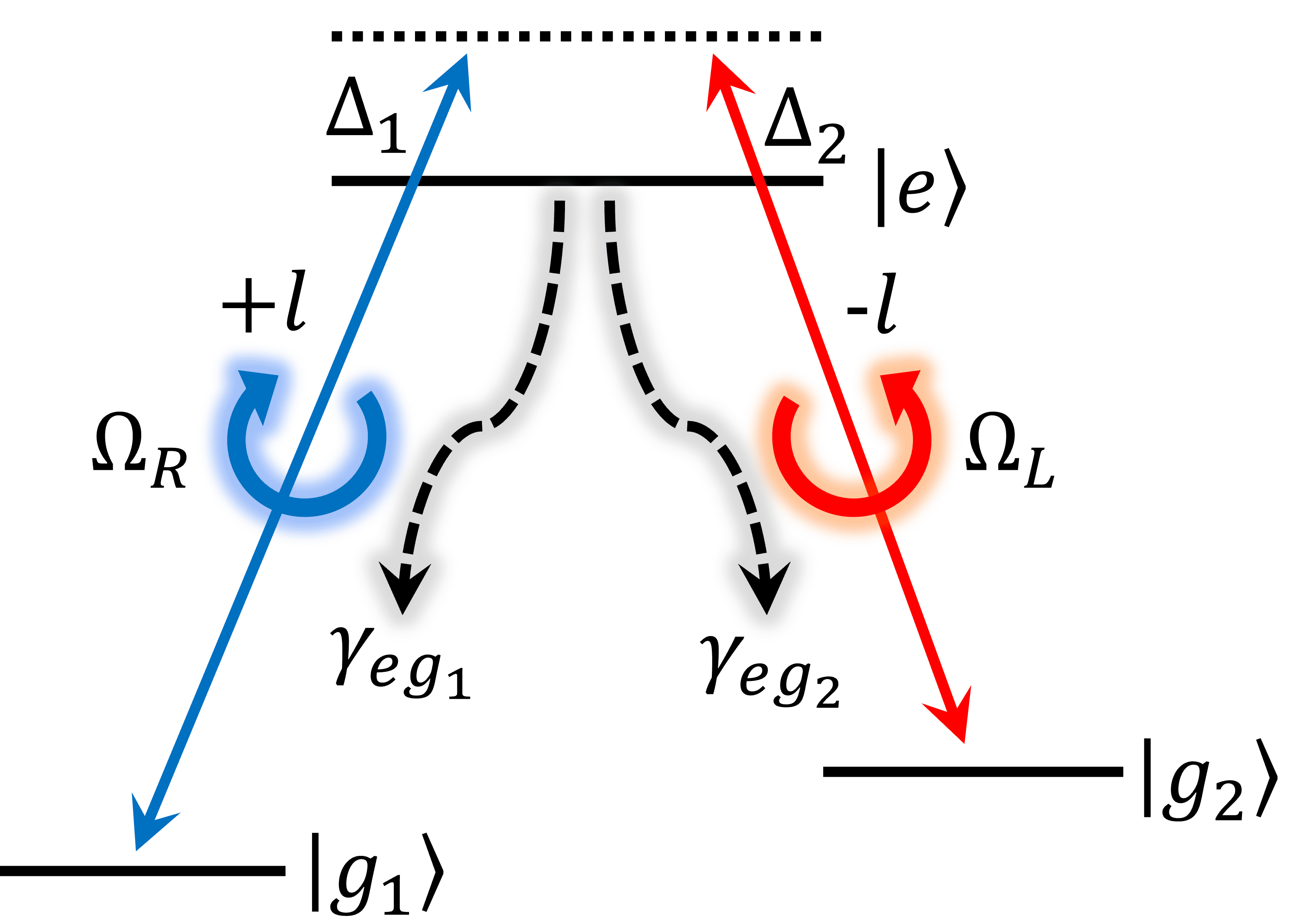}
    \caption{Schematic of the three-level atomic system forming a $\Lambda$-type configuration. The excited state $\ket{e}$ is coupled to the ground states $\ket{g_1}$ and $\ket{g_2}$ by right- and left-circularly polarized light fields with Rabi frequencies $\Omega_R$ and $\Omega_L$, respectively.}
    \label{Lambda Configuration}
\end{figure}

We consider the propagation of a weak optical vector vortex beam through a coherently prepared atomic medium in $\Lambda$ configuration. The vector beam consists pulse pairs of two circularly polarized components with right- and left-handed polarization, each carrying opposite OAM charges. The total electric field can be written as $\vec{E} = E_L \hat{e}_L + E_R \hat{e}_R$, where $E_L$ and $E_R$ are the slowly varying envelopes associated with the left- and right-circularly polarized unit vectors $\hat{e}_L$ and $\hat{e}_R$, respectively. The corresponding Rabi frequencies of the right- and left-circular components are denoted by $\Omega_R$ and $\Omega_L$, and are related to the electric field via $\Omega_{L(R)}=\vec{d}_{L(R)}\cdot \vec{E}_{L(R)}/\hbar$, where $\vec{d}_{L(R)}$ is the dipole moment of the atomic transition corresponding to the electric field component $\vec{E}_{(L,R)}$ \cite{scully.book1997}. The atomic medium consists of two ground states, $\ket{g_1}$ and $\ket{g_2}$, and one excited state, $\ket{e}$. The right-circularly polarized component with Rabi-frequency $\Omega_R$ couples the transition $\ket{g_1} \leftrightarrow \ket{e}$, while the left-circularly polarized component with $\Omega_L$ couples the transition $\ket{g_2} \leftrightarrow \ket{e}$, thus forming a $\Lambda$-type configuration, as illustrated in Fig.~\ref{Lambda Configuration}. Under the dipole approximation, the Hamiltonian describing the atom–light interaction can be expressed as
\begin{subequations}
\begin{equation}
    H = H_0 + H_I, \label{eq.1a}
\end{equation}
\begin{equation}
    H_0 = \hbar\omega_{g_1}\ket{g_1}\bra{g_1} + \hbar\omega_{g_2}\ket{g_2}\bra{g_2} + \hbar\omega_{e}\ket{e}\bra{e}, \label{eq.1b}
\end{equation}
\begin{equation}
    H_I = \hbar\Omega_R^*e^{-i\omega_Rt}\ket{e}\bra{g_1}+\hbar\Omega_L^*e^{-i\omega_Lt}\ket{e}\bra{g_2} + H.c., \label{eq.1c}
\end{equation}
\label{eq.1}
\end{subequations}
where $H_0$ denotes the atomic Hamiltonian expressed in the basis of the bare states, and
, and $H_I$ is the dipole interaction Hamiltonian.

To express the Hamiltonian in a time-independent form under an appropriate rotating-wave frame, we introduce the following unitary transformation
\begin{equation}
    U = e^{i(\omega_e - \omega_R)t\ket{g_1}\bra{g_1}+i(\omega_e - \omega_L)t\ket{g_2}\bra{g_2}+i\omega_et\ket{e}\bra{e}}, \label{eq.2}
\end{equation}
under which the Hamiltonian transforms according to
\begin{equation}
    H' = i\hbar\frac{\partial U}{\partial t}U^\dagger + UHU^\dagger.\label{eq.3}
\end{equation}
Applying this transformation yields
\begin{multline}
        H' =-\hbar\Delta_1\ket{g_1}\bra{g_1} - \hbar\Delta_2\ket{g_2}\bra{g_2} \\ 
        + \hbar\Omega_R\ket{g_1}\bra{e} + \hbar\Omega_L\ket{g_2}\bra{e} + H.c. , \label{eq.4}
\end{multline}
where the detunings of the optical fields are defined as
\begin{subequations}
    \begin{equation}
        \Delta_1  = \hbar\omega_{e g_1} - \hbar\omega_R; \quad \omega_{e g_1} = \omega_e - \omega_{g_1}, \label{eq.5a}
    \end{equation}
    \begin{equation}
        \Delta_2 = \hbar\omega_{e g_2} - \hbar\omega_L; \quad \omega_{e g_2} = \omega_e - \omega_{g_2}. \label{eq.5b} 
    \end{equation}
    \label{eq.5}
\end{subequations}

\subsection{\label{Equation of Motion}Equations of Motion}
We assume the system relaxes solely via spontaneous emission from the excited state $\ket{e}$ to the ground states $\ket{g_1}$ and $\ket{g_2}$, with decay rates $\gamma_{e g_1}$ and $\gamma_{e g_2}$, respectively. The system dynamics are described by the evolution of the atomic density matrix via the Liouville–von Neumann equation
\begin{equation}
    \dot{\rho} = -\frac{i}{\hbar}\left[H,\rho\right] - \frac{1}{2}\left\{L,\rho \right\}. \label{eq.6}
\end{equation}
The decay operator $L$ can be written as 
\begin{equation}
    L_{ij} = \delta_{ij}\gamma_j, \label{eq.7}
\end{equation}
with $\delta_{ij}$ the Kronecker delta and $\gamma_j$ the total decay rate associated with state $\ket{j}$. Substituting the time-independent Hamiltonian from Eq.~(\ref{eq.4}) and the decay operator from Eq.~(\ref{eq.7}) into Eq.~(\ref{eq.6}) yields the following set of optical Bloch equations for the coherence terms
\begin{subequations}
    \begin{multline}
        \dot{\rho}_{g_1 e} = i (\Delta_1+i\gamma_{e g_1}) \rho_{g_1 e} - i\Omega_R \left( \rho_{ee} - \rho_{g_1 g_1}\right) \\ 
        + i \Omega_L \rho_{g_1 g_2} - \gamma_{e g_1} \rho_{g_1 e}, \label{eq.8a} 
    \end{multline}
    \begin{multline}
        \dot{\rho}_{g_2 e} = i (\Delta_2+i\gamma_{e g_1}) \rho_{g_2 e} - i\Omega_L \left( \rho_{ee} - \rho_{g_2 g_2}\right)\\
        + i \Omega_R \rho_{g_2 g_1} - \gamma_{e g_2} \rho_{g_2 e}, \label{eq.8b} 
    \end{multline}
    \label{eq.8}
\end{subequations}
where the transition decay rates are related to the state decay rates by $\gamma_{e g_1}=\frac{\gamma_e + \gamma_{g_1}}{2}$ and $\gamma_{e g_2}=\frac{\gamma_e + \gamma_{g_2}}{2}$.

We assume that the atoms are initially prepared in a coherent superposition of the ground states, so that the atomic state before interaction with the laser fields is
\begin{equation}
    \ket{\Psi(0)} = c_1 \ket{g_1} + c_2 \ket{g_2}. \label{eq.9}
\end{equation}
We consider both laser fields to be weak, satisfying $|\Omega_R|,|\Omega_L| \ll \gamma_{e g_1},\gamma_{e g_2}$. Therefore, the steady state solution for $\rho_{ij}$ can be found using perturbative expansion up to the first order combined with the atoms initial state, giving $\rho_{ee}\approx 0$, $\rho_{g_1 g_1}\approx |c_1|^2$, $\rho_{g_2 g_2}\approx |c_2|^2$, and $\rho_{g_1 g_2}\approx c_1 c_2^*$. This leads to
\begin{subequations}
    \begin{equation}
        \rho_{g_1 e} = -\frac{|c_1|^2 \Omega_R + c_1 c_2^*\Omega_L}{\Delta_1 + i\gamma_{e g_1}}, \label{eq.10a}
    \end{equation}
    \begin{equation}
        \rho_{g_2 e} = -\frac{c_1^* c_2\Omega_R + |c_2|^2 \Omega_L}{\Delta_2 + i\gamma_{e g_2}}. \label{eq.10b}
    \end{equation}
    \label{eq.10}
\end{subequations}
Validity of the first-order approximation requires $|\rho_{g_1 e}|, |\rho_{g_2 e}| \ll 1$. Beyond this regime, in the strong-coupling limit, the ground-state populations $\rho_{g_1 g_1}$ and $\rho_{g_2 g_2}$ evolve significantly during propagation. In the weak-probe regime considered here, however, the excited-state population $\rho_{ee}$ does not appear at first order in the perturbative expansion. Contributions from $\rho_{ee}$ enter only at second order and higher, where they lead to optical pumping and consequently to a depletion of the initially prepared ground-state coherence defined in Eq.~(\ref{eq.9}). In that case, population excited by the probe does not fully return to the original superposition but instead decays incoherently.

These higher-order effects become relevant only in the nonlinear regime when the probe intensities are appreciable. In contrast, our analysis is restricted to the linear regime, where the weak probe ensures that the initial coherence remains effectively unchanged during propagation. Moreover, we treat the probe fields as pulses of finite duration rather than continuous waves. As long as the pulse duration exceeds the optical coherence formation time---set by the spontaneous decay rate---the atomic response remains in the quasi-steady-state regime and the prepared ground-state coherence is preserved. For instance, in $^{87}\mathrm{Rb}$ vapor, the decay rate is typically $\gamma \sim 10~\mathrm{MHz}$, corresponding to a lifetime of $\sim 100~\mathrm{ns}$. Thus, probe pulses longer than $100~\mathrm{ns}$ (readily achievable experimentally) ensure that the atoms adiabatically follow the instantaneous steady-state optical coherence.

\subsection{\label{Equation of Motion}Maxwell–Bloch Equations and Propagation of Vector Vortices}
To study the propagation of optical vector vortices through the atomic medium, we employ the coupled Maxwell–Bloch equations (MBEs) for both circularly polarized components of the field. Within the paraxial and slowly-varying envelope approximations, the MBEs for the right- and left-circularly polarized components $\Omega_R$ and $\Omega_L$ read
\begin{subequations}
    \begin{equation}
        \frac{\partial \Omega_R}{\partial z} + c^{-1} \frac{\partial \Omega_R}{\partial t} = i \frac{\alpha_1 \gamma_{e g_1}}{2L}\rho_{g_1 e}, \label{eq.11a}
    \end{equation}
    \begin{equation}
        \frac{\partial \Omega_L}{\partial z} + c^{-1} \frac{\partial \Omega_L}{\partial t} = i \frac{\alpha_2 \gamma_{e g_2}}{2L}\rho_{g_2 e}, \label{eq.11a}
    \end{equation}
    \label{eq.11}
\end{subequations}
where, $\alpha_1$ and $\alpha_2$ are the optical depth of the medium with length $L$, and $c$ is the vacuum speed of light.

We have assumed that the transverse derivative of fields $\nabla^2_\perp\Omega_R$ and $\nabla^2_\perp\Omega_L$ that corresponds to the diffraction effect in the linear propagation regime can be ignored in Eq.(\ref{eq.11}). A similar treatment was performed in \cite{Hamid.PRA2019,Kudriasov.OE2025} for different types of atom-light configuration, which was assumed to be a valid approximation if the phase change due to the diffraction effect is lower than $\pi$. If we estimate the transverse derivative components as $\nabla^2_\perp \Omega \sim w^{-2} \Omega$, and the temporal variation of the fields as $\frac{\partial\Omega}{\partial t} \sim \frac{c}{L}\Omega$, then the phase shift due to the diffraction is given by $\frac{L}{2kw^2}$. Here, $k=\frac{2\pi}{\lambda}$ is the wave number of the beams with the wavelength $\lambda$, and $w$ corresponds to transverse dimension of the fields, which is the characteristic width for Gaussian beam without vortex, or the width of the vortex core for beam carrying vortex. Therefore, the condition of which the diffraction effect can be neglected is $\frac{L \lambda}{w^2} \ll \pi$, which can be satisfied by choosing the appropriate sample medium length for a certain range of wavelength $\lambda$.

The steady-state solutions for the atomic coherences in Eq.(\ref{eq.10}) when inserted into the MBEs of Eq.(\ref{eq.11}) produce a set of coupled differential equations governing the propagation of the beams in the atomic medium:  
\begin{subequations}
    \begin{equation}
        \frac{\partial \Omega_R}{\partial z} = -i \beta_1 \left(|c_1|^2 \Omega_R+ c_1 c_2^* \Omega_L\right), \label{eq.12a}
    \end{equation}
    \begin{equation}
        \frac{\partial \Omega_L}{\partial z} = -i \beta_2 \left(c_1^*c_2 \Omega_R+ |c_2|^2 \Omega_L\right), \label{eq.12b}
    \end{equation}
    \label{eq.12}
\end{subequations}
where $\beta_j$ (with $j=1,2$) is defined as
\begin{equation}
    \beta_j = \frac{\alpha_j \gamma_{e g_j}}{2L \left(\Delta_j + i\gamma_{e g_j} \right)}. \label{eq.13}
\end{equation}

The analytical solution of Eq.(\ref{eq.12}) can be obtained by specifying the initial conditions at the entrance of the medium $z=0$. Let $\Omega_R(0)=\Omega_{R,0}$ and $\Omega_L(0)=\Omega_{L,0}$ denote the initial values of the fields at the medium entrance. The field solutions of Eq.(\ref{eq.12}) take the form:
\begin{subequations}
\begin{equation}
    \Omega_R(z) = \frac{q_1}{X} \Omega_{R,0} + \frac{q_2}{X} \Omega_{L,0}, \label{eq.14a}
\end{equation}
\begin{equation}
    \Omega_L(z) = \frac{q_4}{X} \Omega_{R,0} + \frac{q_3}{X} \Omega_{L,0}, \label{eq.14b}
\end{equation}
\label{eq.14}
\end{subequations}
where the coefficients $q_1,q_2,q_3,q_4$ and $X$ are defined as
\begin{subequations}  
\begin{align}
    q_1 &= \beta_2 |c_2|^2 + \beta_1 |c_1|^2 e^{-iX z}, \label{eq.15a}\\
    q_2 &= \beta_1 c_1 c_2^* \left(e^{-iX z} - 1 \right), \label{eq.15b}\\
   q_3 &= \beta_1 |c_1|^2 + \beta_2 |c_2|^2 e^{-iX z}, \label{eq.15c} \\ 
   q_4 &= \beta_2 c_1^* c_2 \left(e^{-iX z} - 1 \right), \label{eq.15d} \\
   X &= \beta_1 |c_1|^2 + \beta_2 |c_2|^2. \label{eq.15e}
\end{align}
\label{eq.15}
\end{subequations}

The evolution of $\Omega_R$ and $\Omega_L$ components during propagation through the atomic medium can be analyzed using Eq.(\ref{eq.14}). Together with Eq.(\ref{eq.15}), these equations show that both of light beams will experience loss during propagation due to the imaginary part of $X$. However, the loss will only present at the medium entrance before EIT regime is established for both beams. A more detailed discussion about the absorption is presented in Section \ref{Results and Discussion}. Moreover, the solutions were derived assuming that both beams are initially present at the medium entrance, but these equations can also be used to study the generation of the secondary beam and OAM transfer when only one beam is present at the beginning \cite{Hamid.PRA2019}. This can be implemented simply by setting one of the input fields to zero at the entrance. In what follows, we will employ these solutions to study various cases of propagation of vector beams with different polarization states and OAM charges.

\section{\label{Results and Discussion}Results and Discussion}
In this section, we first analyze the characteristic propagation distance of the beams inside the atomic medium, which defines the typical length within which absorption is dominant. Furthermore, we investigate the absorption patterns and propagation dynamics of vector vortices in the coherently prepared $\Lambda$ medium, highlighting how the interplay of SAM and OAM influences the evolution of the beams.

\subsection{\label{Characteristic Propagation Distance} Characteristic Propagation Distance}
The evolution of the beams described by Eq.~(\ref{eq.14}) is governed by the medium parameters contained in the coefficients $q_1, q_2, q_3, q_4$ and $X$ defined in Eq.(\ref{eq.15}). The exponential terms in Eq.(\ref{eq.15a})-(\ref{eq.15d}) depend on the generally complex coefficient $X$. Separating $X$ into its real and imaginary parts, the exponential terms can be expressed as
\begin{equation}
    e^{-iXz} = e^{-i\rm{Re}\left[ X \right]z}e^{\rm{Im}\left[ X \right]z}. \label{eq.16}
\end{equation}
In Eq.(\ref{eq.16}), the real part of $X$ governs the oscillatory behavior of $\Omega_R$ and $\Omega_L$ during propagation through the atomic medium, while the imaginary part accounts for exponential damping.

For simplicity, we consider $\alpha_1=\alpha_2=\alpha$, and $\gamma_{e g_1}=\gamma_{e g_2}=\gamma$, giving $\beta=\frac{\alpha\gamma}{2L(\Delta_j+i\gamma)}$. We also assume that both laser fields have the same detuning $\Delta_1=\Delta_2=\Delta$, which leads to $\beta_1=\beta_2=X$ according to Eq.(\ref{eq.13}). In the resonant case ($\Delta=0$), we define the absorption length as $L_{abs}=\frac{L}{\alpha}$ representing the distance over which the fields experience absorption in the medium. For propagation distances much larger than the absorption length ($z \gg L_{abs}$), the exponential terms in Eq.(\ref{eq.15a}-\ref{eq.15d}) vanish, since the imaginary part of $X$ dominates, leading to field extinction. In this limit, Eq.(\ref{eq.14}) and Eq.(\ref{eq.15}) simplify to
\begin{subequations}
\begin{align}
    \Omega_R(z\gg L_{abs}) &= |c_2|^2 \Omega_{R,0} - c_1 c_2^*\Omega_{L,0}, \label{eq.17a} \\ 
    \Omega_L(z\gg L_{abs}) &= -c_1^*c_2\Omega_{R,0} + |c_1|^2\Omega_{L,0}. \label{eq.17b}
\end{align}
\label{eq.17}    
\end{subequations}

As obtained in Eq.(\ref{eq.17}), the effect of attenuation disappears after the beams have propagated a sufficiently large distance, and both fields reach a stationary condition where their amplitudes remain constant. In this regime, the EIT is established, allowing the beams to propagate without loss within the medium—similar to free-space propagation \cite{Marangos.RMP2005}. However, the establishment of EIT also depends on the initial atomic coherence coefficients $c_1$ and $c_2$, as well as on the input beam parameters at the medium entrance, $\Omega_{R,0}$ and $\Omega_{L,0}$. For the specific case where $c_1=c_2$ and $\Omega_{R,0}=\Omega_{L,0}$ (equal amplitudes and equal OAM charges), one gets complete extinction of both $\Omega_R$ and $\Omega_L$ from Eq.(\ref{eq.17}) at large propagation distances. In this case, the medium does not become transparent, and therefore, the beam amplitude is continuously attenuated throughout propagation until extinction is reached.
Consequently, the realization of EIT can be actively tuned by adjusting the relative amplitudes and phases of the optical fields and the initial atomic coherence. 

 In the off-resonant case, the propagation distance required for lossless propagation increases, and the fields exhibit damped oscillations because $X$ is no longer purely imaginary. The characteristic attenuation distance $z_c$ in this regime is defined as
\begin{equation}
    z_c = -\frac{1}{2}\frac{1}{\rm{Im}\left[X\right]}, \label{eq.18} 
\end{equation}
where the factor of $\frac{1}{2}$ serves as a normalization term ensuring that $z_c = L_{abs}$ under resonance conditions ($\Delta = 0$). From the definition of $X$ in Eq.(\ref{eq.15e}), it follows that the characteristic distance $z_c$ is independent of the initial choice of atomic coherence coefficients $c_1$ and $c_2$ when $\beta_1 = \beta_2 = \beta$, since in this case $X=\beta$. Consequently, $z_c$ depends solely on the medium parameters, namely, the optical depth $\alpha$, decay rate $\gamma$, and the probe detuning $\Delta$, through their relation to $\beta$. 

\begin{figure}[h!]
    \centering
    \includegraphics[width=1\linewidth]{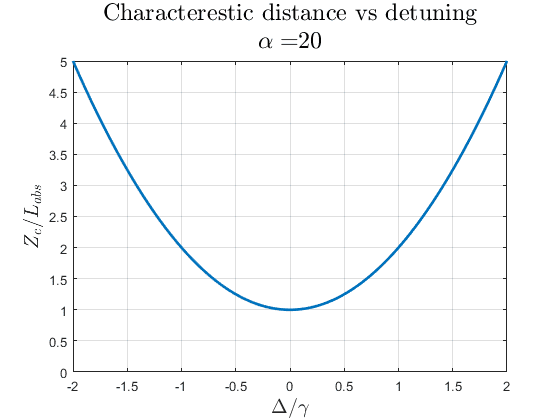}
    \caption{Characteristic distance $z_c$ as a function of detuning $\Delta$. The curve exhibits a parabolic dependence on $\Delta$. The detuning is normalized to the decay rate $\gamma$, and the characteristic distance is normalized to the absorption length $L_{abs}$ such that, at resonance, $z_c = L_{abs}$.}
    \label{Characterstic Distance Plot}
\end{figure}

In Fig.~\ref{Characterstic Distance Plot}, we present the characteristic distance $z_c$ as a function of the detuning $\Delta$. The figure shows that as the system moves further away from resonance, the characteristic distance $z_c$ increases in a parabolic manner. This behavior indicates that the propagation length required to establish EIT becomes longer at larger detunings. Once EIT is established, however, the beams propagate through the medium without absorption and eventually reach a stationary state for propagation distances $z \gg z_c$. Therefore, it is desired to examine the propagation dynamics and absorption behavior up to the distance where this stationary regime is achieved.

\subsection{\label{Absorption Pattern}Azimuthal Absorption and EIT Patterns}
Let us now consider that the beam entering the atomic medium is an optical vector vortex, formed by two circularly polarized components with opposite OAM charges \cite{Kudriasov.OE2025}. Their coherent superposition gives rise to the full vector vortex structure. To simplify the discussion, we assume that the polarization components are represented by Laguerre–Gaussian (LG) modes with the lowest radial order:

\begin{subequations}
 \begin{align}
     \Omega_R(0) &= \Omega_{R,0} = |\Omega_{R,0}|\mathrm{LG}^R=\epsilon_R A(r)e^{il\phi}, \label{eq.19a} \\
    \Omega_L(0) &= \Omega_{L,0}=|\Omega_{L,0}|\mathrm{LG}^L=\epsilon_L A(r) e^{-il\phi}, \label{eq.19b}
 \end{align}
\label{eq.19} 
\end{subequations}
where $|\Omega_{R,0}|$ and $|\Omega_{L,0}|$ are the Rabi amplitudes for the right-handed and left-handed beam, respectively, and $A(r)$ describes the transverse amplitude profile of the beam, which is given by
\begin{equation}
    A(r) = \left(\frac{r}{w}\right)^{|l|}e^{-\frac{r^2}{w^2}}, \label{eq.20}
\end{equation}
with $w$ being the beam waist. The relative amplitudes $\epsilon_R$ and $\epsilon_L$ can be expressed as 
\begin{subequations}
\begin{align}
    \epsilon_L &= \varepsilon \cos{(\theta)}, \label{eq.21a} \\
    \epsilon_R &= \varepsilon \sin{(\theta)} e^{i\psi}, \label{eq.21b}
\end{align}   
\label{eq.21}
\end{subequations}
where $\theta$ is a tuning parameter that controls the relative strength between the left-handed amplitude and the right-handed amplitude, and $\psi$ is the phase difference between the two LG modes at the entrance of the medium, while $\varepsilon$ is an arbitrary scaling constant. Consequently, the resulting vortex beam at the entrance of the medium can be written as
\begin{multline}
    \vec{E}(r,\phi,z=0) = E_L(0)\hat{e}_L + E_R(0) \hat{e}_R\\ 
    = \varepsilon\cos{(\theta)}LG^L \hat{e}_L + \varepsilon \sin{(\theta)}e^{i\psi}LG^R \hat{e}_R
    . \label{eq.22}
\end{multline}

When the vector vortex beam interacts with the $\Lambda$-type atomic medium, the evolution of its right and left components is governed by Eq.(\ref{eq.14}). By substituting the input vortex fields from Eq.(\ref{eq.19}) into Eq.(\ref{eq.14}), we obtain
\begin{subequations}
\begin{align}
    \Omega_R(z) &= \frac{q_1}{X} |\Omega_{R,0}|\mathrm{LG^R} + \frac{q_2}{X} |\Omega_{L,0}|\mathrm{LG^L}, \label{eq.23a}
    \\
    \Omega_L(z) &= \frac{q_4}{X} |\Omega_{R,0}|\mathrm{LG^R} + \frac{q_3}{X} |\Omega_{L,0}|\mathrm{LG^L}. \label{eq.23b}
\end{align}
\label{eq.23}
\end{subequations}
Equation (\ref{eq.23}) describes the coupled evolution of both field components along the propagation direction $z$, showing how the individual vortex components interact within the medium and laying the groundwork for later analysis of the resulting vector vortex behavior.

Before analyzing the propagation dynamics of the resulting vector vortex beam, let us first examine the absorption behavior of the individual vortex probe components at different propagation distances within the medium. This can be achieved by substituting Eq.~(\ref{eq.23}) into the steady-state coherence expression given in Eq.~(\ref{eq.10}). For the analysis of absorption properties, we focus on a specific case where the initial atomic coherence is symmetric, i.e., $c_1 = c_2 = \frac{1}{\sqrt{2}}$ and the system is on resonance $\Delta=0$, leading to $\beta=-\frac{i\alpha}{2L}$. Under this condition, and according to Eq.~(\ref{eq.15}), the coefficients $q_1$, $q_2$, $q_3$, and $q_4$ reduce to 
\begin{subequations}
\begin{align}
   \tilde{q}_1 &= \tilde{q}_3 = -\frac{i \alpha}{4 L} \left( 1 + e^{-\frac{\alpha}{2 L} z} \right), \label{eq.24a} \\
    \tilde{q}_2 &= \tilde{q}_4 = -\frac{i \alpha}{4 L} \left( -1 + e^{-\frac{\alpha}{2 L} z} \right). \label{eq.24b}
\end{align}
\label{eq.24}
\end{subequations}

Assuming $\theta=\pi/4$ and $\psi=0$, such that $|\Omega_{R,0}|=|\Omega_{L,0}|=\varepsilon A(r)$, from Eq.(\ref{eq.23}) and (\ref{eq.24}), the propagation of fields follow
\begin{subequations}
\begin{align}
    \Omega_R(z) & = \varepsilon A(r) \left(e^{-\frac{\alpha}{2L} z} \cos{(l\phi)} + i \sin{(l\phi)} \right), \label{eq.25a}
    \\
    \Omega_L(z) & = \varepsilon A(r) \left(e^{-\frac{\alpha}{2L} z} \cos{(l\phi)} - i \sin{(l\phi)} \right), \label{eq.25b}
\end{align}
\label{eq.25}
\end{subequations}
where $\varepsilon A(r)$ represents the radial amplitude distribution of the LG modes described in Eq.(\ref{eq.20}), which is identical for both the right- and left-handed components. Inserting Eq.(\ref{eq.25}) into Eq.(\ref{eq.10}), yields the steady-state atomic coherences, varying radially, azimuthally, and longitudinally (in resonance $\Delta=0$), which take the form 
\begin{equation}
    \rho_{g_1 e}(r,\phi,z) = \rho_{g_2 e}(r,\phi,z) = \frac{\varepsilon A(r)}{\gamma}\cos{(l\phi)}\left(ie^{-\frac{\alpha}{2L} z}\right). \label{eq.26}
\end{equation}

\begin{figure}
    \centering
    \includegraphics[width=1\linewidth]{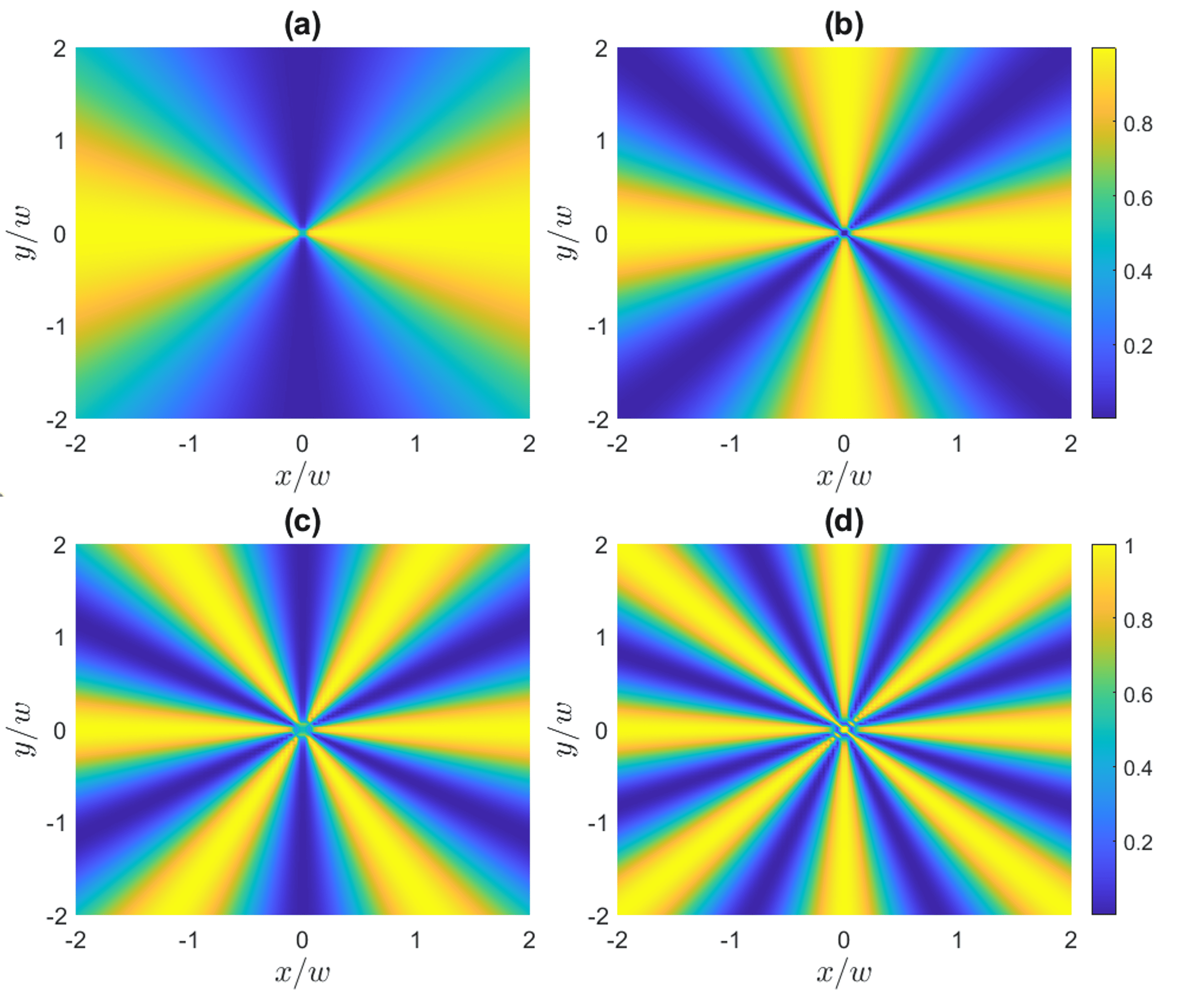}
    \caption{Absorption pattern of the right-handed beam on resonance $\Delta=0$ for different topological charge $|l|$ at $z/L_{abs}=0$: (a) $|l|=1$, (b) $|l|=2$, (c) $|l|=3$, and (d) $|l|=4$. Other parameters are $\alpha=20$, $c_1=c_2=\frac{1}{\sqrt{2}}$}
    \label{Absorption Different L}
\end{figure}

The steady-state coherences in Eq.~(\ref{eq.26}) reflect the spatial profile and topological charge $l$ of the input vortex beams as they propagate through the medium. In particular, the $\cos{(l\phi)}$ factor preserves the azimuthal symmetry of the coherence, indicating that the medium inherits the topological structure of the input fields.
While these terms describe how the internal atomic states vary spatially within the medium, they do not directly reveal the absorption experienced by the fields. To quantify absorption, one must consider the medium susceptibility $\chi$, which in the linear regime is proportional to the atomic coherence. For a transition $\ket{i}\rightarrow\ket{j}$ driven by a light field with Rabi frequency $\Omega_k$, the corresponding linear susceptibility is given by \cite{Marangos.RMP2005}:
\begin{equation}
    \chi_{ij} \sim \frac{\rho_{ij}}{\Omega_k}, \label{eq.27}
\end{equation}
where $i,j \in \{g_1,g_2,e\}$ and $k\in\{R,L\}$.

Combining Eqs.~(\ref{eq.25})–(\ref{eq.27}) yields the expressions for the medium susceptibilities experienced by the right- and left-handed beams:
\begin{subequations}
\begin{align}
\chi_{R} &\sim \frac{\rho_{g_1 e}}{\Omega_R} = \frac{\cos{(l\phi)}}{\gamma}\frac{ie^{-\frac{\alpha}{2L}z}}{e^{-\frac{\alpha}{2L}z} \cos{(l\phi)} + i \sin{(l\phi)}} , \label{eq.28a} 
\\
\chi_{L} &\sim \frac{\rho_{g_2 e}}{\Omega_L} = \frac{\cos{(l\phi)}}{\gamma}\frac{ie^{-\frac{\alpha}{2L}z}}{e^{-\frac{\alpha}{2L}z} \cos{(l\phi)} - i \sin{(l\phi)}} , \label{eq.28b} 
\end{align}
\label{eq.28}
\end{subequations}

\begin{figure}[h!]
    \centering
    \includegraphics[width=1\linewidth]{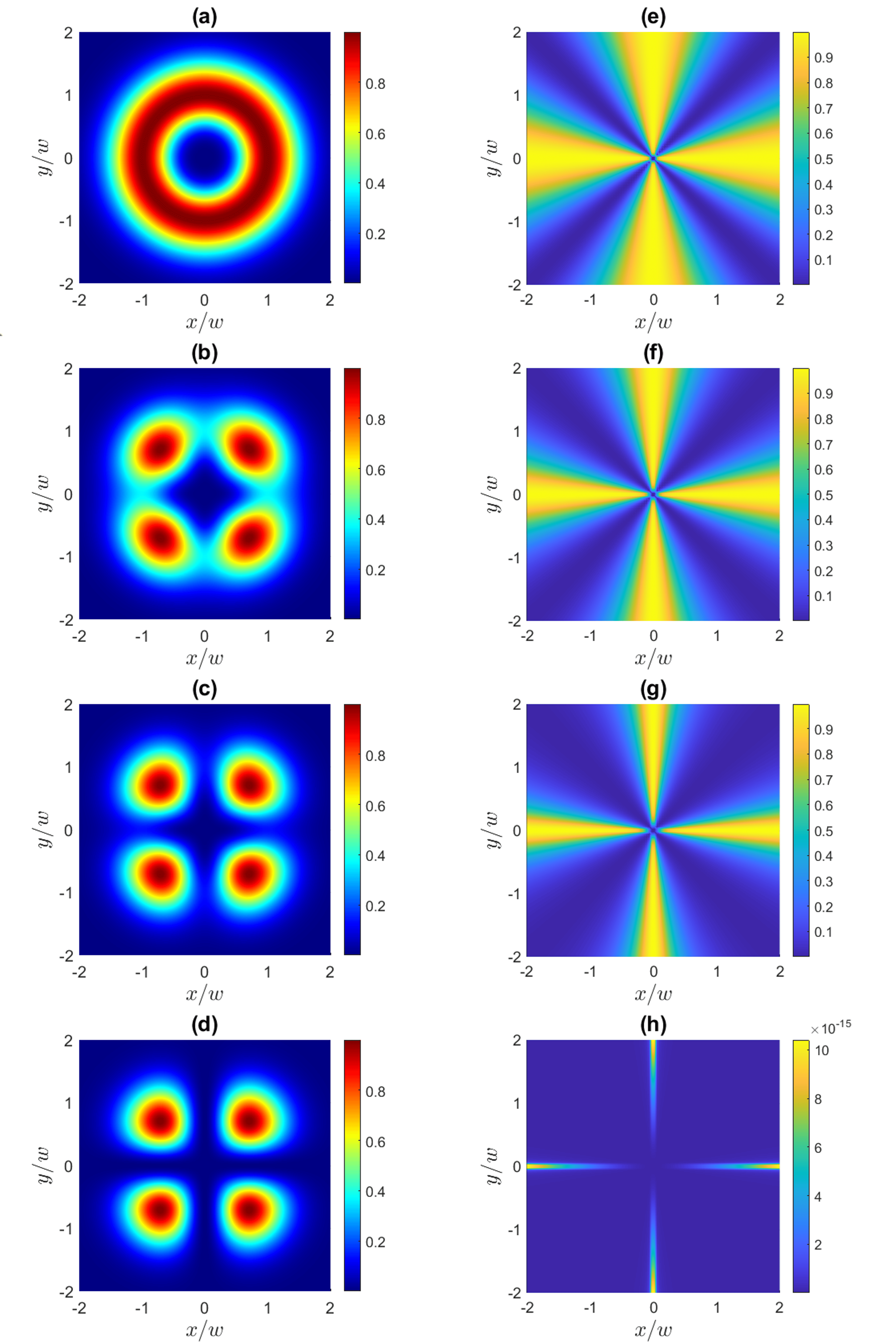}
    \caption{Intensity distributions (a)–(d) and absorption patterns (e)–(h) of the right-handed beam with $|l|=2$ at different propagation distances: (a),(e) $z/L_{abs}=0$; (b),(f) $z/L_{abs}=0.5$; (c),(g) $z/L_{abs}=1$; and (d),(h) $z/L_{abs}=20$. The other parameters are the same as those used in Fig.\ref{Absorption Different L}.}
    \label{Absorption Different z}
\end{figure}

We use the analytical expressions in Eq.~(\ref{eq.28}) to plot the absorption patterns for beams with various OAM charges $|l|$ and propagation distances $z$. The propagation distance is normalized to the absorption length $L_{abs}$, allowing the results to be scaled for any medium length $L$. Figure~\ref{Absorption Different L} illustrates the absorption pattern of the right-handed beam for different topological charges. For a beam with charge $|l|$, the absorption pattern exhibits $2|l|$ transparency windows (regions of reduced absorption) in the azimuthal plane, consistent with the $\cos(l\phi)$ dependence in Eq.~(\ref{eq.26}).

Figure~\ref{Absorption Different z} shows the evolution of the absorption pattern and the corresponding intensity distribution for a right-handed beam with $|l|=2$ at various propagation distances. As the beam propagates deeper into the medium, absorption regions in the azimuthal plane shrink and their magnitude decreases, resulting in broader transparency windows. Correspondingly, the beam intensity evolves from a doughnut-shaped profile at the entrance ($z/L_{abs}=0$) into a petal-like structure with four intensity maxima as it propagates. These intensity peaks coincide with the transparency regions in the absorption map, where the field experiences minimal losses. Furthermore, even though the transparency regions expand with propagation, the overall topological structure of the absorption pattern remains conserved, and the number of transparency windows ($2|l|$) does not change with distance.

\begin{figure*}[!t]
\centering
\includegraphics[width=\textwidth]{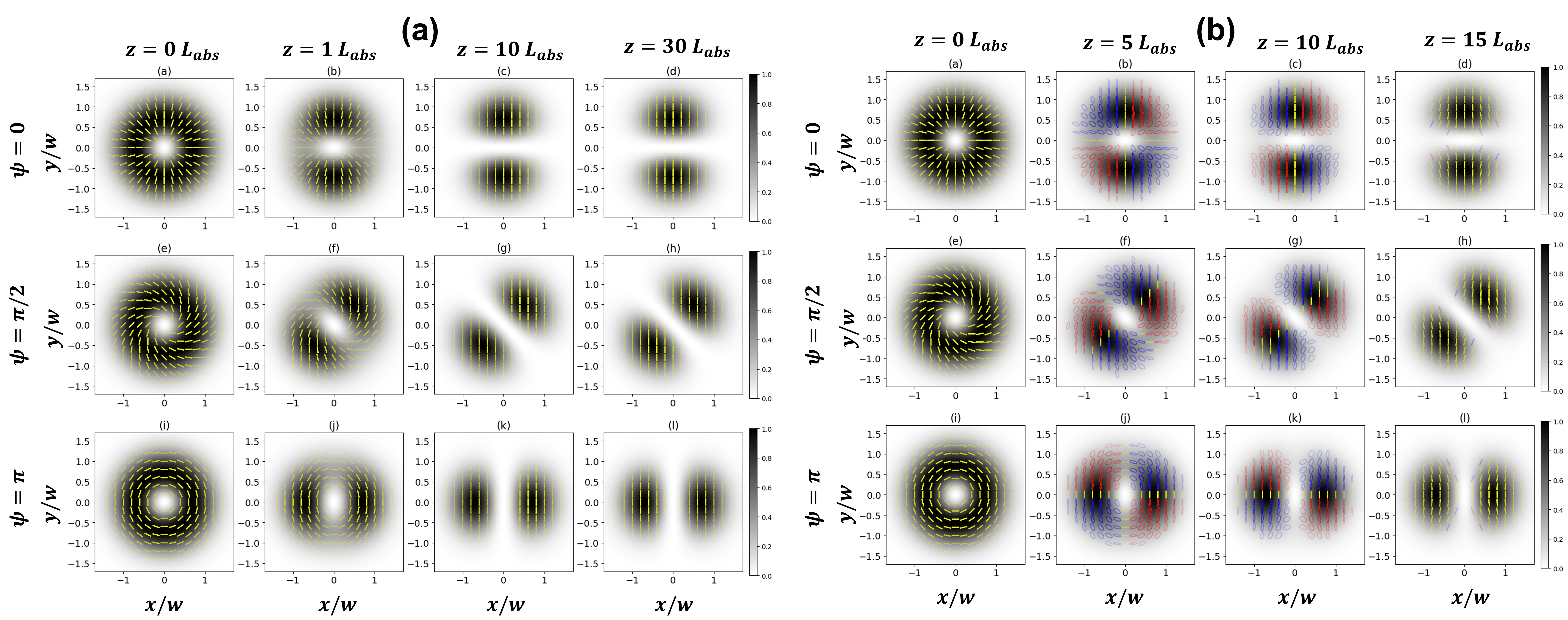}
\caption{Intensity and polarization state distributions in the transverse plane of a vector vortex beam carrying topological charge $|l|=1$ and relative amplitude $\theta=\pi/4$, corresponding to equal left- and right-handed field strengths ($|E_L|=|E_R|$ since $\cos{\theta}=\sin{\theta}$). The transverse coordinates $(x, y)$ are normalized to the beam waist $w$, and the propagation distance $z$ is normalized to the absorption length $L_{abs}$. The medium parameters are $c_1=c_2=1/\sqrt{2}$ and $\alpha=20$. Two detuning cases are shown: (a) resonant ($\Delta=0$) and (b) off-resonant ($\Delta=2\gamma$). The first, second, and third rows correspond to relative phases $\psi=0$, $\pi/2$, and $\pi$, respectively. Darker rings or lobes indicate higher intensity regions. The yellow lines denote linearly polarized states, while red and blue ellipses represent left- and right-handed circular polarizations, respectively.}
\label{c1c2_equal_amplitude_equal}
\end{figure*}

\subsection{\label{Propagation Dynamics}Vector Vortices Propagation Dynamics}
In this section, we analyze in detail the propagation dynamics of the initial vector vortex beam with its components carrying opposite orbital angular momenta, satisfying $l_R = -l_L = 1$. The initial vector vortex beam, formed by a superposition of counter-rotating LG modes, evolves into a more complex field configuration with spatially varying polarization as it propagates through the atomic medium.
The total electric field of the resulting vector vortex beam at any propagation position $z$ can be written as $\vec{E}(r,\phi,z) = E_L(z)\hat{e}_L + E_R(z)\hat{e}_R$, where $E_R(z)$ and $E_L(z)$ are obtained from the analytical solutions in Eq.~(\ref{eq.23}), allowing for arbitrary initial beam amplitudes and general atomic superposition coefficients $c_1$ and $c_2$. Explicitly, these field components are expressed as
\begin{subequations}
\begin{align}
    E_R &= \varepsilon \left(\sin{(\theta)}e^{i\psi}\mathrm{LG^R}\frac{q_1}{X}+ \cos{(\theta)}\mathrm{LG^L}\frac{q_2}{X}\right), \label{eq.29a}
    \\
    E_L &=\varepsilon \left(\sin{(\theta)}e^{i\psi}\mathrm{LG^R}\frac{q_4}{X}+ \cos{(\theta)}\mathrm{LG^L}\frac{q_3}{X} \right). \label{eq.29b}
\end{align}
\label{eq.29}
\end{subequations}
The spatial evolution of the beam polarization distribution during propagation can then be characterized using the Stokes parameters in the circular polarization basis \cite{Tarak.OE2022} 
\begin{subequations}
\begin{align}
    S_0 &= |E_R|^2 + |E_L|^2, \label{eq.30a} \\
    S_1 &= 2\mathrm{Re}(E_R^*E_L), \label{eq.30b}\\
    S_2 &=2\mathrm{Im}(E_R^*E_L), \label{eq.30c}\\
    S_3 &=|E_R|^2-|E_L|^2. \label{eq.30d}
\end{align}
\label{eq.30}
\end{subequations}

The Stokes parameters defined in Eq.~(\ref{eq.30}) are used to describe the beam’s ellipticity $\zeta=\frac{1}{2}\sin^{-1}{(\frac{S_3}{S_0})}$ and orientation angle $\xi=\frac{1}{2}\tan^{-1}{(\frac{S_2}{S_1})}$, which define the polarization states of the propagating beams inside the atomic medium. Three distinct polarization cases can be identified based on the ellipticity: linear polarization for $\zeta=0$, circular polarization for $\zeta=\pm\pi/4$, and elliptical polarization for $0 < |\zeta| <\pi/4$. On the other hand, the orientation angle $\xi$ determines the local rotation of the polarization at each point in the transverse plane, and the rotation after propagation over a distance $z$ is given by $\Delta\xi=\xi(z)-\xi(0)$.

Using the Stokes parameters defined in Eq.~(\ref{eq.30}) together with the field components in Eq.~(\ref{eq.29}), we can analyze the intensity distribution and polarization dynamics of the beams inside the atomic medium. We consider both the resonant and off-resonant cases. For the resonant case, the propagation distances are chosen as $z/L_{abs} = 0, 1, 10, 30$ to illustrate the initial changes in polarization and intensity at short distances, and then the evolution in the regime where EIT is fully established, i.e., at larger propagation lengths. In the off-resonant case, shorter distances of $z/L_{abs} = 0, 1, 10, 15$ are sufficient, because the SAM in this case is more sensitive to propagation due to the non-zero detuning, which induces oscillations between polarization states.

\begin{figure*}[!t]
\centering
\includegraphics[width=1\textwidth]{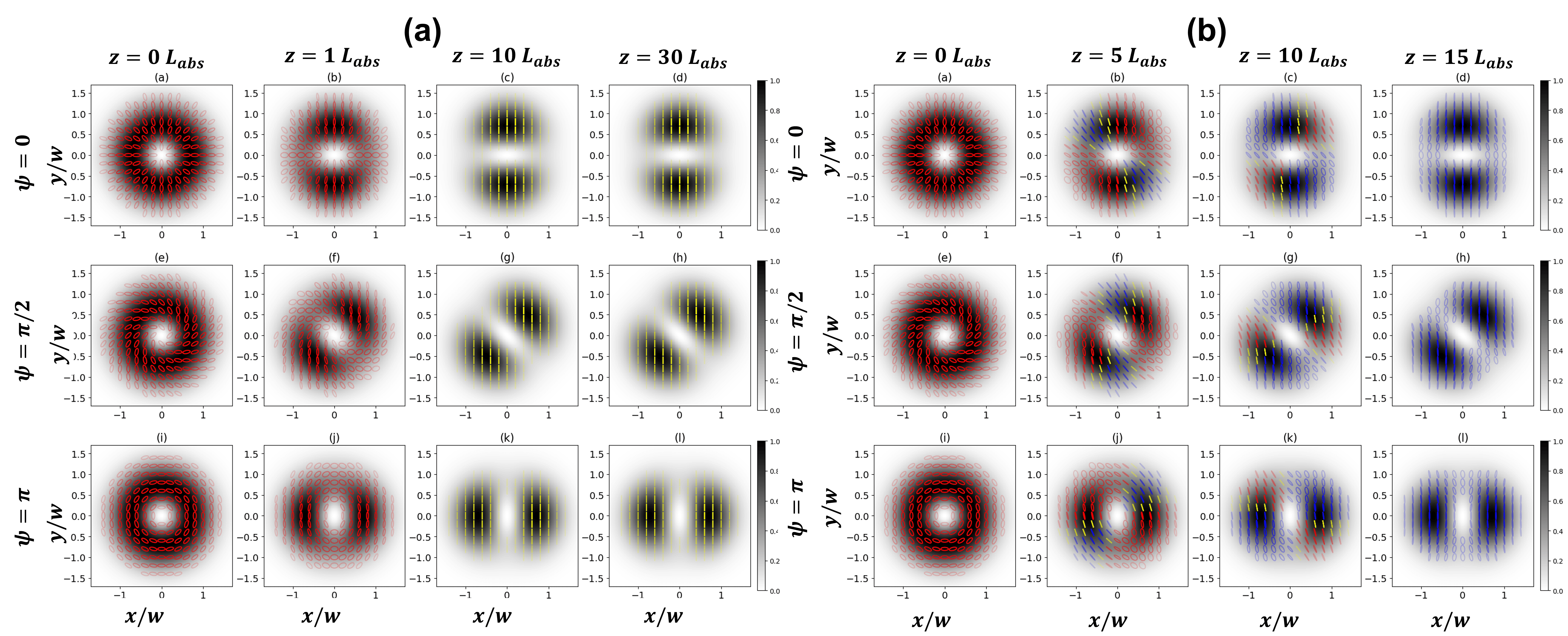}
\caption{ Intensity and polarization state distributions in the transverse plane of a vector vortex beam carrying topological
 charge $|l|=1$ and relative amplitude $\theta=\pi/8$, such that the initial left-handed beam dominates over the right-handed beam $|E_L(0)|>|E_R(0)|$. The transverse coordinates $(x, y)$ are normalized to the beam radius $w$, and the propagation distance $z$ is normalized to the absorption length $L_{abs}$. The medium parameters are $c_1=c_2=1/\sqrt{2}$, and $\alpha=20$. Two detuning cases
 are shown: (a) resonant $\Delta=0$, and (b) non-resonant $\Delta=2\gamma$. The first, second, and third rows correspond to relative
 phases $\psi=0,\pi/2$, and $\pi$, respectively. Darker rings or lobes indicate higher intensity regions. The yellow lines denote linearly
 polarized states, while red and blue ellipses represent left- and right-handed circular polarizations, respectively.}
\label{c1c2_equal_amplitude_nonequal}
\end{figure*}

We first consider the case where the atomic superposition coefficients have equal probability amplitudes, $c_1=c_2=\frac{1}{\sqrt{2}}$. Figure \ref{c1c2_equal_amplitude_equal}(a) corresponds to the resonant case $\Delta=0$, while Fig.\ref{c1c2_equal_amplitude_equal}(b) corresponds to the non-resonant case with $\Delta=2\gamma$, with equal initial beam amplitudes for both cases ($|\Omega_{R,0}|=|\Omega_{L,0}|$). At the entrance of the medium $z=0$, both cases display a doughnut-shaped intensity profile accompanied by a uniform linear polarization across the transverse plane. Three distinct polarization textures are presented for each case, radial  ($\psi=0$), spiral ($\psi=\pi/2$), and azimuthal ($\psi=\pi$), determined by the choice of the relative phase $\psi$. The linear polarization at $z=0$ arises from the equal amplitudes of the right- and left-handed components, whose superposition produces a linearly polarized state.

Under resonance (Fig.\ref{c1c2_equal_amplitude_equal}(a)), the intensity profile evolves into a petal-shaped structure with two distinct lobes at larger propagation distances. These lobes correspond to the regions where the beam intensity inside the medium is not attenuated, representing the transparency windows observed in Fig.\ref{Absorption Different L} for $|l|=1$. As shown for the cases $\psi=\pi/2$ and $\psi=\pi$, the transparency windows, and consequently the lobes, rotate by an angle of $\psi/2$. Furthermore, the linear polarization is maintained throughout the propagation, indicating that the strength of the right- and left-handed components remains balanced. 


Contrary to the resonant case, the non-resonant case (Fig.~\ref{c1c2_equal_amplitude_equal}(b)) exhibits an oscillatory evolution of the polarization state. Here, the polarization transitions from a purely linear state into a mixed regime where linear, left-circular (red ellipses), and right-circular (blue ellipses) polarizations coexist at $z/L_{abs}=5$ and $10$, before reverting to linear polarization at $z/L_{abs}=15$. These oscillations arise from the position-dependent phase difference between the right- and left-handed field components, introduced by the exponential phase factor $e^{-iXz}$ in the coupling coefficients defined in Eq.~(\ref{eq.15}). This phase modulation leads to a periodic exchange of SAM between the optical field and the atomic medium, manifesting as cyclic conversion between linear and circular polarization states.

\begin{figure*}[!t]
\centering
\includegraphics[width=\textwidth]{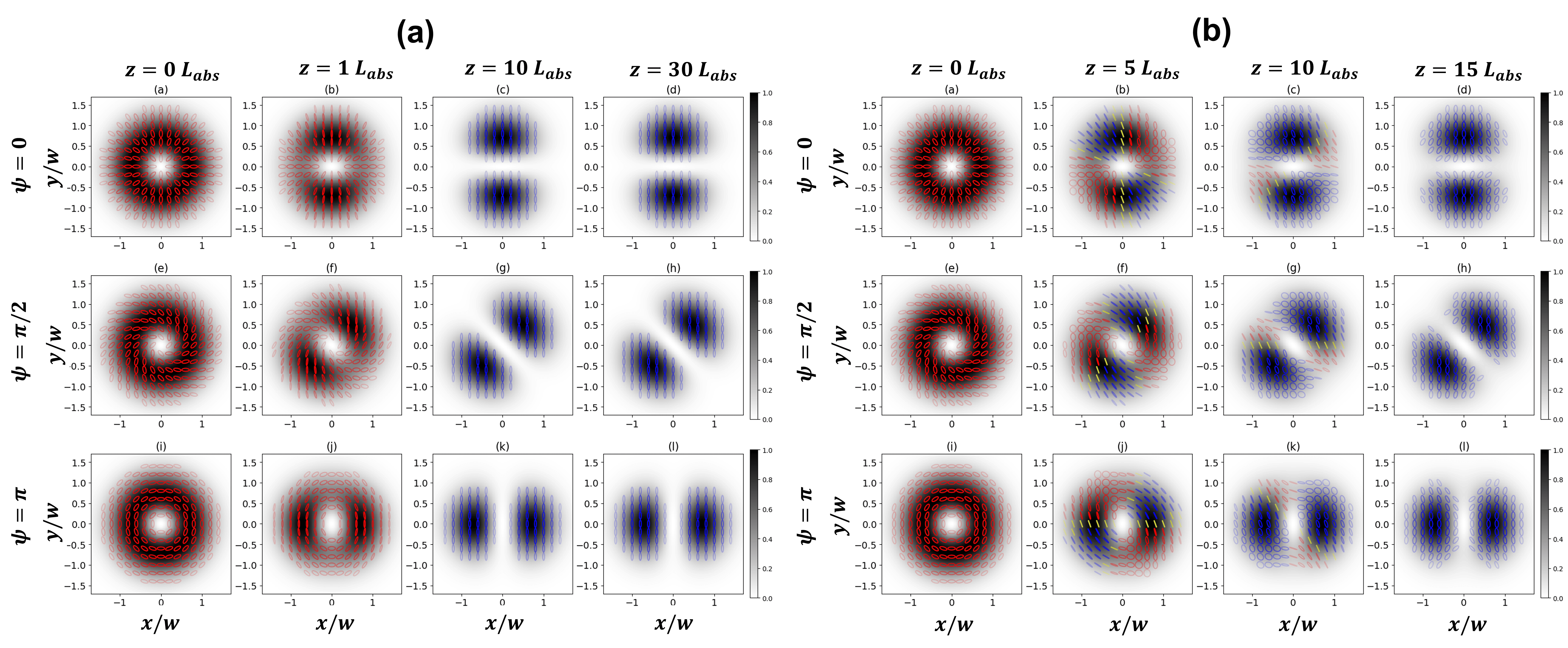}
\caption{Intensity and polarization stated distributions in the transverse plane of a vector vortex beam carrying topological
 charge $|l|=1$ and relative amplitude $\theta=\pi/8$,  such that the initial left-handed beam dominates over the right-handed beam $|E_L(0)|>|E_R(0)|$. The transverse coordinates $(x,y)$ are normalized to the beam radius $w$, and the propagation distance is normalized to the absorption length $L_{abs}$. The medium parameters are $c_1<c_2$ with $c_1=1/2$, and $\alpha=20$. Two detuning cases are shown: (a) resonant $\Delta=0$, and (b) non-resonant $\Delta=2\gamma$. The first, second, and third rows correspond to relative phase $\psi=0,\pi/2$, and $\pi$, respectively. Darker rings or lobes indicate higher intensity regions. The yellow lines denote linearly polarized states, while red and blue ellipses represent left- and right-handed circular polarizations, respectively.}
\label{c1c2_nonequal_amplitude_nonequal}
\end{figure*}

In Fig.~\ref{c1c2_equal_amplitude_nonequal}, we investigate the case with relative amplitude $\theta = \pi/8$, such that the initial left-handed beam at the medium entrance ($z=0$) is stronger than the right-handed beam ($|E_L(0)| > |E_R(0)|$). In this configuration, a complete polarization transition from one state to another is observed for both the resonant and non-resonant cases. In the resonant case [Fig.~\ref{c1c2_equal_amplitude_nonequal}(a)], the beam initially exhibits a left-handed circular polarization across the transverse plane, corresponding to dominant left-handed SAM. As the beam propagates inside the medium, the polarization gradually evolves toward a linear state at $z/L_{abs} = 10$ and $30$. This transition from left-handed circular to linear polarization reflects a redistribution of SAM between the left and right components mediated by the coherent atom–field interaction.  Once the left- and right-handed components reach equal amplitudes ($|E_L(z)| = |E_R(z)|$), the beam acquires a linear polarization state. After entering the EIT regime at larger distances, the energy transfer process ceases, and the resulting linear polarization is maintained throughout the propagation.
For the non-resonant case [Fig.~\ref{c1c2_equal_amplitude_nonequal}(b)], the beam exhibits polarization oscillations at intermediate propagation distances $z/L_{abs} = 5$ and $10$, where linear, left-circular, and right-circular polarizations coexist across the transverse plane, similar to the non-resonant case in Fig.~\ref{c1c2_equal_amplitude_equal}(b). In contrast to the resonant scenario, where the beam eventually reaches a stable linear polarization, the non-resonant propagation leads to a complete polarization flip from left-handed circular to right-handed circular polarization. This SAM reversal arises from the detuning, which modifies the energy transfer between the circular components, resulting in a final state dominated by the right-handed component ($|E_R|>|E_L|$).

Now, we consider the case where the initial atomic coherences are unequal, $c_1 \neq c_2$. In the coherently prepared state given by Eq.~(\ref{eq.9}), the probability amplitudes $c_1$ and $c_2$ satisfy $|c_1|^2 + |c_2|^2 = 1$ to conserve total probability. A relative phase $\varphi_c$ between $c_1$ and $c_2$ can be introduced, such that $c_2 = |c_2| e^{i\varphi_c}$, while still satisfying probability conservation. For illustration, we consider $c_1 = 1/2$ and $\varphi_c = 0$, so that $c_1 < c_2$. All other parameters and variations are kept the same as in Fig.~\ref{c1c2_equal_amplitude_nonequal}, where the initial beam at the medium entrance exhibits dominant left-handed circular polarization.

Figure~\ref{c1c2_nonequal_amplitude_nonequal} illustrates the case $c_1 < c_2$, where the left-handed beam initially dominates at the medium entrance ($z=0$). In both the resonant [Fig.~\ref{c1c2_nonequal_amplitude_nonequal}(a)] and non-resonant [Fig.~\ref{c1c2_nonequal_amplitude_nonequal}(b)] scenarios, the polarization transitions from left-handed to right-handed. Unlike the case with equal atomic coherences ($c_1 = c_2 = 1/\sqrt{2}$), where polarization flip occurs only in the presence of detuning, unequal atomic coherences induce the flip even under resonance. The resonant and non-resonant cases differ in transition dynamics: in the non-resonant case, oscillatory behavior appears at intermediate propagation distances due to the real part of $X$, whereas in the resonant case $X$ remains purely imaginary, producing a monotonic transition. Additionally, the relative amplitudes of $c_1$ and $c_2$ determine the contributions of left- and right-handed components through the coefficients $q_1, q_2, q_3,$ and $q_4$ in Eq.~(\ref{eq.23}), shaping the SAM redistribution and the evolving polarization pattern.

Previously, a similar case of polarization transition and rotation has been investigated in the four-level tripod configuration under the influence of an external magnetic field \cite{Tarak.OE2022,Kudriasov.OE2025}. The applied magnetic field introduces anisotropy inside the medium, giving rise to circular birefringence and dichroism. In such systems, the anisotropy originates from the dependence of the medium susceptibilities on the right- and left-handed circular components that drive the ground-to-excited-state transitions under the action of the magnetic field. The magnetic field breaks the degeneracy of these transitions and produces unequal detunings for the two circular components, making the right-handed field effectively red-detuned and the left-handed field blue-detuned, or vice versa. The degree of anisotropy is therefore fully determined by the applied magnetic field. In the absence of this external field, the medium becomes isotropic, so the polarization dynamics vanish. 

In contrast, the anisotropy responsible for the polarization dynamics in the present three-level $\Lambda$ system arises purely from the atomic coherence. The initial ground-state superposition introduced in Eq.~(\ref{eq.9}) acts as an internal anisotropy, breaking the isotropy of the medium through quantum coherence even though no external magnetic field is applied. The nonzero ground-state coherence term $c_{1}c_{2}^{*}$ couples the right-handed and left-handed circular components in the propagation equations, as shown in Eq.~(\ref{eq.14}), resulting in complex susceptibilities $\chi_{R}$ and $\chi_{L}$ given in Eq.~(\ref{eq.28}), which are no longer identical even in the simplest symmetric case where the input beams have equal amplitudes, the initial superposition satisfies $c_{1}=c_{2}=1/\sqrt{2}$ and $\Delta=0$. In the more general situation where the input amplitudes differ, the initial populations satisfy $c_{1}\neq c_{2}$ and $\Delta \neq 0$, the anisotropy of susceptibilities becomes even more pronounced. This coherence-induced anisotropy acts back on the propagating vector beam, giving rise to spin–orbit coupling of light, which manifests as polarization exchange.

 \section{Concluding Remarks}

To summarize, we have investigated the propagation dynamics of optical vector vortex pulse pairs composed of orthogonal RCP and LCP components carrying opposite OAM as they interact with an ensemble of atomic systems in a $\Lambda$ configuration. The atoms were initially prepared in a phaseonium state, i.e., a coherent superposition of the two ground levels. Under the dipole and rotating-wave approximations and within the weak-interaction regime, we obtained the steady-state solution for the optical coherences. These coherences were then used to derive analytical solutions for the propagation of the vortex pairs through the medium via the Maxwell–Bloch equations (MBEs) in the linear regime, where diffraction was neglected. The MBEs solutions enabled us to analyze the characteristic distance $z_c$ required for the establishment of EIT, the formation of topology-dependent transparency windows, and the polarization evolution of the vector vortex pairs.

For vortex pairs with opposite OAM charges, a transparency region (EIT regime) emerges after the beams propagate over distances much larger than the characteristic scale, $z \gg z_c$. At propagation distances comparable to $z_c$, the EIT condition is not yet fully established and the beams experience absorption in the azimuthal plane. The characteristic distance was shown to depend on the detuning $\Delta$: larger detuning requires longer propagation to achieve transparency due to the oscillatory response of the medium under non-resonant excitation.

We demonstrated that the topology of the vortex pairs—specified by their OAM charge $|l|$—is imprinted onto the medium through the spatial structure of the induced coherence. This imprinting manifests as the formation of transparency windows with $2|l|$ azimuthal symmetry. These transparency windows reshape the transmitted vortex intensity, transforming the initial doughnut-like profile into a petal-like structure whose lobes correspond to the transparent azimuthal regions. During propagation, the transparency windows gradually broaden until EIT is fully established, while robustly preserving their $2|l|$ topological symmetry.

Importantly, the OAM-structured atomic coherence induces an internal anisotropy and circular dichroism in the medium. This coherence-induced anisotropy acts back on the propagating vector vortex beam and mediates an effective optical spin-orbit coupling, resulting in an exchange of SAM between the polarization components and a rotation of the overall polarization state. The strength of this spin-orbit interaction can be tuned by adjusting the initial ground-state populations and the relative amplitudes of the vortex components. In particular, when the initial ground states are unequally populated ($c_1 \neq c_2$), the induced anisotropy becomes more pronounced and manifests as a clear polarization conversion, including the transformation of LCP into RCP light.

Lastly, the phaseonium system we have investigated offers the advantage to control optical vector vortices without the need of additional external magnetic field. The coherent superposition of the ground states is possible to realized experimentally using stimulated Raman adiabatic passage (STIRAP) \cite{bergmann2017} that allows the selective transfer of the population into the target quantum states without suffering additional loss from stimulated emission. In the three level $\Lambda$ system, STIRAP is done in practice by controlling the pulse delay between the pump and probe pulses such that adiabatic evolution occurs that transfers the population of the ground state $\ket{g_1}$ to the ground state $\ket{g_2}$ via the dark state route (superposition of $\ket{g_1}$ and $\ket{g_2}$) without populating the excited state $\ket{e}$. Whereas, the possible physical elements to realize the three-level $\Lambda$ system can be implemented in $^{87}\mathrm{Rb}$ atomic vapours \cite{lukin2001}. In this atomic vapours system, the Zeeman sub-level $\ket{^5P_{1/2}, F=1,m_f=0}$ can be used as the excited state $\ket{e}$, while the ground state $\ket{g_1}$ and $\ket{g_2}$ can be implemented in the Zeeman sub-level $\ket{^5S_{1/2}, F=2,m_f=+2}$ and $\ket{^5S_{1/2}, F=2,m_f=0}$.  

\begin{acknowledgments}
This project has received funding from the Research Council of Lithuania (LMTLT), agreement No. S-ITP-24-6. D.P.P. gratefully acknowledges Dr. Viačeslav Kudriašov for the technical help on simulation of vortices, and gratefully acknowledges the support of the Erasmus+: Erasmus Mundus programme of the European Union under
Convention n° 101128124 — EUROPHOTONICS — ERASMUS-EDU-2023-PEX-EMJMMOB. 
\end{acknowledgments}

\bibliography{apssamp}

\end{document}